\documentclass[conference]{IEEEtran}
\IEEEoverridecommandlockouts
\usepackage{cite}
\usepackage{amsmath,amssymb,amsfonts}
\usepackage{algorithm}
\usepackage{graphicx}
\usepackage{textcomp}
\usepackage{xcolor}
\usepackage{bbm}
\usepackage{url}

\usepackage{cite}
\usepackage{balance}

\usepackage{graphicx}
\usepackage{subcaption}

\ifCLASSINFOpdf
\else
\fi

\DeclareMathAlphabet\mathbfcal{OMS}{cmsy}{b}{n}
\usepackage{algorithm}      
\usepackage{algorithmic}

\usepackage{varwidth}

\begin{document}

\title{
V2X System Architecture Utilizing Hybrid Gaussian Process-based Model Structures}

\author{Hossein Nourkhiz Mahjoub$^*$, Behrad Toghi$^*$, S M Osman Gani$^*$, Yaser P. Fallah$^*$ \\
$^*$Networked Systems Lab \\ Department of Electrical and Computer Engineering, University of Central Florida, Orlando, FL\\
\{hnmahjoub, toghi, smosman.gani\}@knights.ucf.edu, Yaser.Fallah@ucf.edu
        
\thanks{This material is based on work supported in part by the National Science
Foundation under CAREER Grant 1664968 and in part by the Qatar
National Research Fund Project NPRP 8-1531-2-651.}}



\maketitle

\begin{abstract}
Scalable communication is of utmost importance  for reliable dissemination of time-sensitive information in cooperative vehicular ad-hoc networks (VANETs), which is, in turn, an essential prerequisite for the proper operation of the critical cooperative safety applications. The model-based communication (MBC) is a recently-explored scalability solution proposed in the literature, which has shown a promising potential to reduce the channel congestion to a great extent. In this work, based on the MBC notion, a technology-agnostic hybrid model selection policy for Vehicle-to-Everything (V2X) communication is proposed which benefits from the characteristics of the non-parametric Bayesian inference techniques, specifically Gaussian Processes. The results show the effectiveness of the proposed communication architecture on both reducing the required message exchange rate and increasing the remote agent tracking precision.
 
\end{abstract}
\begin{IEEEkeywords}
Vehicular ad-hoc network, scalable V2X communication, model-based communication, non-parametric Bayesian inference, Gaussian processes. 
\end{IEEEkeywords}
\section{Introduction}
In 1999, the Federal Communication Commission (FCC) allocated 75 MHz of spectrum at the 5.9 GHz frequency for the emerging field of Intelligent Transportation Systems (ITS). Different vehicular communication solutions such as Dedicated Short-Range Communication (DSRC) \cite{dsrc:fcc}, \cite{jkenney:dsrcmain}, \cite{ChMod:ITS} and Cellular Vehicle-to-Everything (C-V2X) \cite{3gpp:36885}, \cite{btoghi:vnc}
have been introduced and developed afterwards, aiming at facilitating the establishment of critical cooperative safety applications, e.g., Forward Collision Warning/Avoidance (FCW/A) \cite{FCW:ITS}, Cooperative Adaptive Cruise Control (CACC) \cite{CACC:Shladover}, \cite{CACC:TIV}, and Intersection Management.

The fundamental role of the V2X communications is enabling every vehicle in a Vehicular Ad-hoc NETwork (VANET) to frequently inform the surrounding nodes about its most recent dynamic states. In general, the V2X architecture could be broken down into three main categories, i.e., communication among vehicles (V2V), communication between vehicles and infrastructure (V2I), and communication between vehicles and Vulnerable Road Users (VRUs), e.g. V2P \cite{atahmasbi:v2p}.

The concept of  information sharing among nodes results in a level of {\textit{situational awareness}} for any vehicle/VRU and makes it aware of its surrounding environment, which is super crucial for the cooperative safety applications to function properly. The Society of Automotive Engineers (SAE), as the main vehicular regulatory organization in US, has proposed a specific framework through a set of standards in order to realize the notion of situational awareness in vehicular networks. The content of the Basic Safety Message (BSM), which conveys the situational awareness information, has been specified by SAE J2735 standard\cite{sae:j2735}. 
However, vehicular networks can potentially experience very dense scenarios which result in a congested communication channel and impose severe performance degradation to the network. Therefore, part of another standard by SAE, i.e. SAE system requirement standard or SAE J2945/1 \cite{sae:j2945}, explores different congestion control mechanisms such as BSM transmission power and rate control in order to manage the generated load from the information beaconing and mitigate the congestion imposed on the communication channel. It is noteworthy that the congestion control algorithms defined by SAE J2945/1 standard do not impose any restrictions on the BSM content or size since these parameters are defined through SAE J2735. In the current SAE framework, the message content remains intact for all broadcast packets\footnote{The terms ``BSM" and ``packet'' are sometimes used interchangeably in this paper} and every BSM is filled out with raw information directly captured from CAN-bus or received from GPS, according to the J2735 dictionary.

The congestion control section of the SAE system requirements standard \cite{sae:j2945}, is the current state-of-the-art congestion control solution accepted by the US vehicular research community as well as US automotive industry. This standard has been developed based on several congestion control algorithms proposed in the literature, among which one can refer to \cite{clhuang:ieeenetwork} and \cite{gbansal:limerictvt}. In a nutshell, the rate and power control algorithms defined in this standard allow vehicles to broadcast their messages at the rate of \mbox{$\sim${1.5--10\ Hz}}. and the power in the range of {10 -- 20\ dBm}, based on their individual network performance evaluation .

In order to improve the performance of the architecture proposed by SAE, a new scheme called Model-Based Communication (MBC) has been recently introduced by the author in \cite{yfallah:mbcsyscon} and more investigated in \cite{emoradipari:tiv2017}, \cite{hnmahjoub:vtc}, and, \cite{hnmahjoub:cavs}. The fundamental intention behind the MBC paradigm is utilizing a more flexible content structure for the broadcast packets based on the joint vehicle-driver predictive behavioral models in comparison with the BSM content structure defined by J2735 standard. This paper, utilizing non-parametric Bayesian modeling schemes, proposes a hybrid model structure within the MBC framework and integrates it with the congestion control communication policy proposed in J2945/1 standard. The notion of MBC and our proposed communication policy will be explored in more details in the subsequent sections.

The rest of this paper is organized as follows. In Section II, an overview of the MBC is presented. Section III is devoted to the system-level architecture design of our proposed model-based communication policy. In this section the proposed hybrid model architecture in addition to the details of our model update policy are thoroughly explained. In section IV, the analysis and evaluation results of the proposed method is presented before the concluding remarks and future research directions stated in Section V.
\section{Model-Based Communication Overview}

One of the main catalysts behind the MBC framework is pursuing a new solution perspective to alleviate the network congestion by re-designing the content structure of the broadcast messages. As stated earlier, the currently standardized dictionary set stipulates the core content of the broadcast messages to be directly filled out with the raw vehicle position and dynamic state update data. Therefore, it does not explicitly reflect the inherent characteristics of the maneuver in which the vehicle is currently involved. However, considering these conceptual characteristics while a vehicle generates its messages could be beneficial for optimizing its scheduled transmission moments. These characteristics could be implicitly utilized to determine the moments at which the instant updates are critical and should be transmitted by the maneuvering vehicle, as well as the moments when transmitting a new packet does not worth. More specifically, in some scenarios, such as abrupt and harsh lane changes or hard brakes an instant update is very critical and highly demanded for the other vehicles' safety applications. On the other hand, multiple redundant transmissions by a vehicle are over-occupying the communication channel if the vehicle, for instance, is cruising in a steady state. In the latter case, transmitting consecutive BSMs not only do not provide its neighbors with any higher degree of situational awareness, but also cause more channel congestion, or equivalently an increase in the number of collided packets, which in turn results in the lower level of situational awareness finally achieved by the neighbouring nodes. From this point of view, the MBC scheme could potentially be capable of improving the communication scalability by scheduling the transmission times at more optimized moments, even if its criteria for this scheduling follows the footsteps of the J2945/1 standard.

 The transmission rate calculation mechanism in J2945/1 is basically based on the transmitter estimation of its surrounding network density, in addition to its estimation of the position tracking accuracy which could be achieved by the information included in its last transmitted BSM. The transmitter keeps track of this tracking precision using an Error-Driven communication mechanism. More precisely, at any GPS update after each BSM transmission, the transmitter calculates the difference between the constant-speed coasting of its position derived from the contents of its last transmitted BSM and its current actual position received via GPS. This difference defines the position tracking error of the transmitter location at this time instance for an arbitrary node which has received the latest transmitted packet. Then,  transmitter performs a comparison between this error with a predefined threshold and decides to transmit a new packet if the error exceeds the threshold. Obviously, this mechanism reduces the transmission rate compared to the baseline 10 Hz transmission.

Now if the transmitted message contains a predictive model with high precision for longer prediction time-horizons, predictions made based upon it at the receiver vehicles could less frequently reach the same position tracking error threshold defined in J2975/1 in comparison with the case of constant speed coasting prediction from raw information received via J2735 BSMs. This explanation clarifies the core idea behind the MBC scheme.

The maneuver characteristics, or equivalently driver behavioral models, are themselves functions of different factors such as the driver's personal driving style, his current mental state, the environmental inputs affecting the driver behavior, e.g. road traffic, other vehicles' maneuvers, weather condition, etc. Considering these factors and reflecting them into the contents of the generated packets by any vehicle is the fundamental idea behind the MBC notion. More specifically, the MBC tries to generate a mathematical model based on the available noise-free CAN-bus information at the transmitter side which be able to explain and predict the driver actions in the future. Assuming these models give notable higher prediction accuracy compared to the constant speed coasting scheme, which is the current default method in the standard, then MBC would be able to avoid several redundant information transmissions. Therefore, the MBC has a two-fold advantage; first it can potentially shrink the payload size by extracting an abstract representation of the vehicle's state. In addition, it reduces the transmission rate by enabling the recipient vehicles to predict their neighbors mobility more accurately in farther time horizons ahead. The former could be achieved through various abstraction and dimensionality reduction methods and the latter would be attained through utilizing different supervised learning algorithms. In this work we have explored the latter case, i.e. the MBC effect on the transmission rate compared to the raw information communication, while the reduced packet size effect is part of our future research directions. 

The initial MBC architecture, illustrated by the author in \cite{yfallah:mbcsyscon}, proposes a stochastic hybrid automata modeling scheme and evaluates its performance on a standard FCW algorithm, known as CAMPLinear \cite{camplinear}. Authors in \cite{emoradipari:tiv2017} use hidden Markov models (HMMs) to derive an adaptive stochastic hybrid system (SHS) in order to capture the non-deterministic nature of driving scenarios. Further enhancements in the modeling approach are presented by authors in \cite{hnmahjoub:vtc} and \cite{hnmahjoub:cavs} which include non-parametric Bayesian inference methods such as Gaussian processes (GPs) with linear kernels and hierarchical Dirichlet process-hidden Markov models (HDP-HMMs). Results in \cite{yfallah:mbcsyscon}-\cite{hnmahjoub:cavs} demonstrate the significant improvements in communication rate and tracking accuracy metrics utilizing the MBC approach.

Analysis in our previous works in \cite{hnmahjoub:vtc} and \cite{hnmahjoub:cavs} demonstrate that the highly dynamic and diverse driving behaviors add more complexity to the modeling process. As an illustration, for the case of a vehicle cruising on a highway, the simplistic constant speed (CS) model will provide an excellent prediction capability. On the contrary, if the vehicle is navigating through a Manhattan-grid urban area, the CS model will be totally obsolete. 

\begin{figure}[t]
\centering
\includegraphics[width=.48\textwidth]{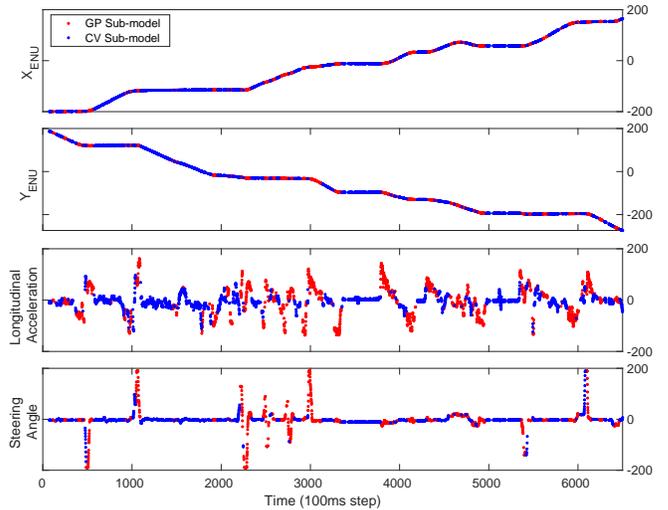}
\caption{Performance of the proposed hybrid modeling scheme in different driving scenarios. Setting a threshold on the tracking error and utilizing a two-state hybrid modeling scheme, compromised of GP (with RBF + Linear kernel) and CV components, this figure (using different colors) shows the moments when each sub-model satisfies the tracking accuracy constraint while the other one fails and exceeds the threshold.}
\label{fig:hybrid_policy_comparison}
\end{figure}

\begin{figure}[htbp]
\centering
\includegraphics[width=.48\textwidth]{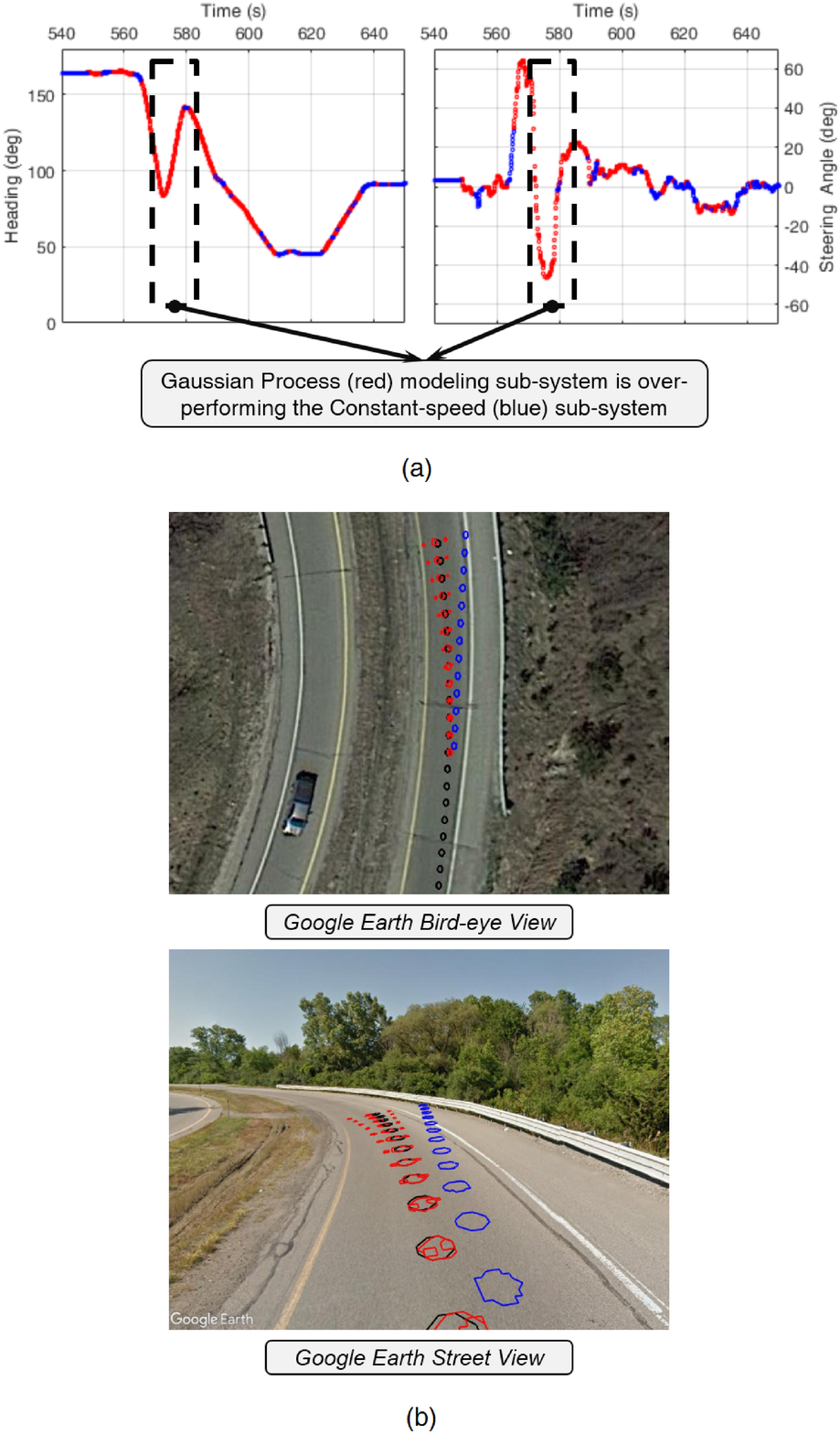}
\caption{Comparison of the prediction precision for GP and CS sub-models. Here, GP outperforms CS because of the maneuver non-linearity. (Courtesy of Google Earth Inc.) }
\label{map_visualization_subplot}
\end{figure}

The above-mentioned phenomena (Figure \ref{fig:hybrid_policy_comparison} and Figure \ref{map_visualization_subplot}) gives an intuition of the core idea in this work; we propose a hybrid modeling architecture which switches between different (here two) modeling sub-systems in order to adapt to the vehicle's dynamic state. Our proposed architecture benefits from a CV modeling sub-system alongside with a GP sub-system with a compound kernel, each of which has shown significant prediction performance in specific scenarios. In addition, since the change points in the high-level driving behaviors on average occur much less frequently compared to the normal message broadcast rates of the state-of-the-art methods in the literature, our hybrid-MBC method gives a conspicuous reduction in required communication rate. The details of our proposed architecture is presented in the next section.

\section{Hybrid GP-based MBC Architecture}
This section provides the details of our proposed communication system architecture composed of the Gaussian process-based modeling block and the error-driven communication framework. In the first subsection, a brief explanation of the Gaussian processes is presented, while combining the hybrid model structures with error-driven communication policy is illustrated in the subsequent subsection.

\subsection{Gaussian Processes: A Fully Data Driven non-parametric Bayesian Modeling Approach}\hfill\\

The record of different vehicle dynamics could be regarded as separate time-series which should be regressed using an appropriate supervised learning method. The regression problem here is equivalent to inferring the characteristics of the unknown target functions which have generated these time-series through their available training sets, which are finite sets of known function output realizations. In this work, following our previous works in \cite{hnmahjoub:vtc} \cite{hnmahjoub:cavs}, a non-parametric Bayesian inference framework is proposed to find an appropriate representation and abstraction of the driver behavior using his observed actions through the recorded time-series of the vehicle dynamics.

In general, the main advantage of any non-parametric inference method is relaxing the function-specific characteristics during the learning process and letting the model complexity to be derived from and adapted to the available training set. In other words, a non-parametric inference method finds the best function representation of the observed data without imposing any prior assumption on the form of the underlying function. Gaussian process (GP), as one of the most powerful non-parametric learning methods, puts the Bayesian prior directly on the function space rather than parameterizing the function and then putting the priors on the parameters space. This trick makes the modeling method capable of capturing different possible patterns which might occasionally be observed in the training data. It is worth mentioning that we use the Gaussian process regression to derive the model of the remote vehicle and its driver as a unique object. The outcome is a set of functions describing the underlying modes which represent the behavior of this object for a notable time ahead.

The formal definition of the Gaussian process is as follows: 
A Gaussian process defines a distribution over function values f(t) at any arbitrary point within the function input range, such that any finite subset of the drawn function values from this distribution form a multivariate Gaussian random vector (have joint Gaussian distribution). \cite{Rasmussen:GP} 

Posterior distribution is inferred by conditioning the problem on a set of noisy observations as the training data.
Gaussian process regression model assumes each observed value as a draw from a normal random variable. Therefore, the set of \textit{m} observations form an \textit{m}-dimensional multivariate normal random vector. This multivariate random vector is defined by a mean vector of length \textit{m} plus an \textit{m}-by-\textit{m} covariance matrix, also called as \textit{kernel} within the Gaussian process context.   

The following equations describe the mathematical representation of GP framework. For more details one can refer to \cite{Rasmussen:GP}. 


\begin{equation}
    f(t) \sim gp (m(t), k(t,t'))
\end{equation}

\begin{equation}
    \{X_i\}_{i=1,2,...,m} = \{f(t_i)\}_{i=1,2,...,m} \sim \mathcal{N}(\overline{\mu},\,\Sigma)\ 
\end{equation}

\begin{equation}
    \overline{\mu} = m(t_i); \ \Sigma_{i,j} = k(t_i, t_j) \ \forall i,j \in \{1,2, ..., m\}
\end{equation}

The kernel matrix defines the correlation between the elements of the marginal distribution. Capturing different patterns is achievable in GP framework by utilizing different types of kernels.  In this work a compound kernel of RBF and linear has shown the best performance in examined specific non-linear scenarios.

\begin{figure}
    \centering
    \begin{subfigure}[b]{0.48\textwidth}
        \includegraphics[width=\textwidth]{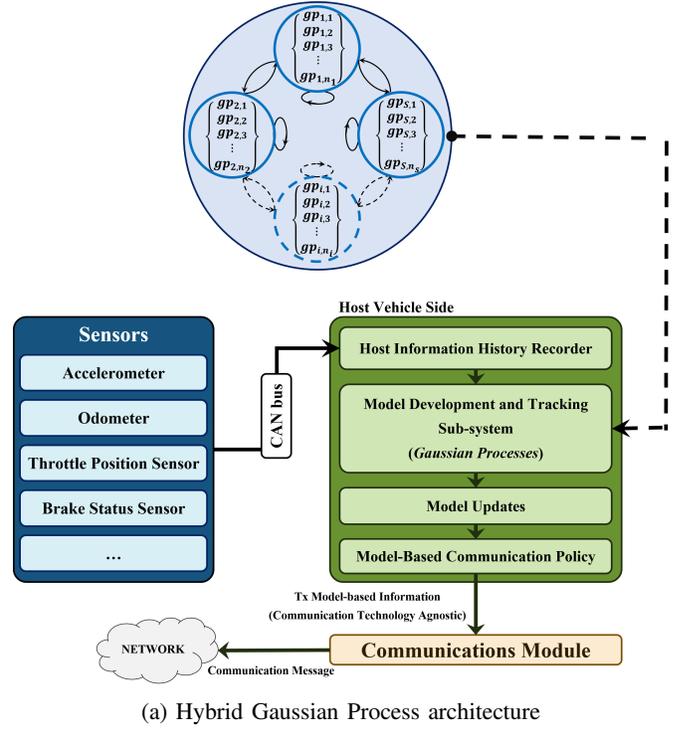}
        \caption{Hybrid Gaussian Process architecture}
        \label{fig:GP_architecture}
    \end{subfigure}
    ~ 
    \begin{subfigure}[b]{0.48\textwidth}
        \includegraphics[width=\textwidth]{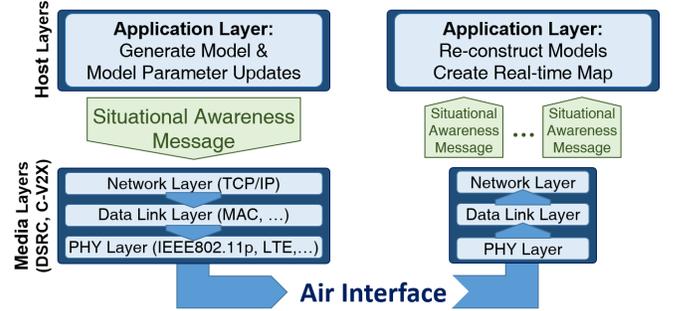}
        \caption{Communication system architecture of a host and remote vehicle}
        \label{fig:communication_architecture}
    \end{subfigure}
    ~ 
    \caption{Systems Design Illustration}\label{fig:system_architecture}
    \label{fig:overall-sys-arch}
\end{figure}


\subsection{An Error-Driven Communication Architecture for Hybrid Model Structures}\hfill\\

As mentioned in the introduction section, inter-vehicle communication in VANETs is our core interest in this work. Assuming a given cooperative vehicular scenario, e.g., a platoon or an intersection, in which each vehicle is equipped with V2X communication devices and interacts with neighboring vehicles, we use the \textit{host} and \textit{remote} vehicle naming conventions as it is common in the vehicular literature. By definition, the host vehicle (HV) receives situational awareness messages from the remote vehicle(s) (RVs) and runs cooperative safety applications locally in order to potentially react to the remote vehicles' actions and maneuvers.
From the networking point-of-view, each network node, i.e., each vehicle, can be modeled as a multi-layer stack. The application layer runs on top of the lower layers which together enable vehicles to communicate over the air-interface.
Considering a RV-HV pair in a network of vehicles, the HV receives multiple situational awareness messages from vehicles in its communication range. As mentioned above, these messages contain dynamical state information which give the HV an insight to create a real-time map of it's surrounding. This map then could potentially be used by the safety applications to avoid collisions or hazardous situations for both HV and RV(s).

Our \textit{communication technology-agnostic} MBC architecture, as illustrated in Figure \ref{fig:overall-sys-arch}, takes place in the application layer and is able to operate independent of the lower network, data-link, and physical layers. Figure \ref{fig:communication_architecture} illustrates the network protocol stack and information flow for an arbitrary HV-RV pair. On the RV side, the Controller Area Network (CAN) bus feeds the application layer with vehicle's local and sensory information. The GP-based MBC module then trains the GP based on the last received information. Afterwards, the MBC module keeps track of the prediction accuracy of the latest learned GP model at any new GPS update and compares it with a certain threshold. Whenever the difference of the latest learned GP prediction and the actual GPS information exceeds this threshold MBC module trains a new GP based on the latest set of sensory inputs. This procedure results in generating a new situational awareness messages which carry the last updated abstract model of the vehicle's state. Lower layers schedule and broadcast the message over the air-interface, i.e., communication channel. The corresponding HV node receives situational awareness messages from all vehicles in its communication range. The MBC module in HV side reconstructs the state of the neighboring vehicles and creates a real-time predictive map of the surrounding nodes.

\begin{figure}[b]
\centering
\includegraphics[width=.48\textwidth]{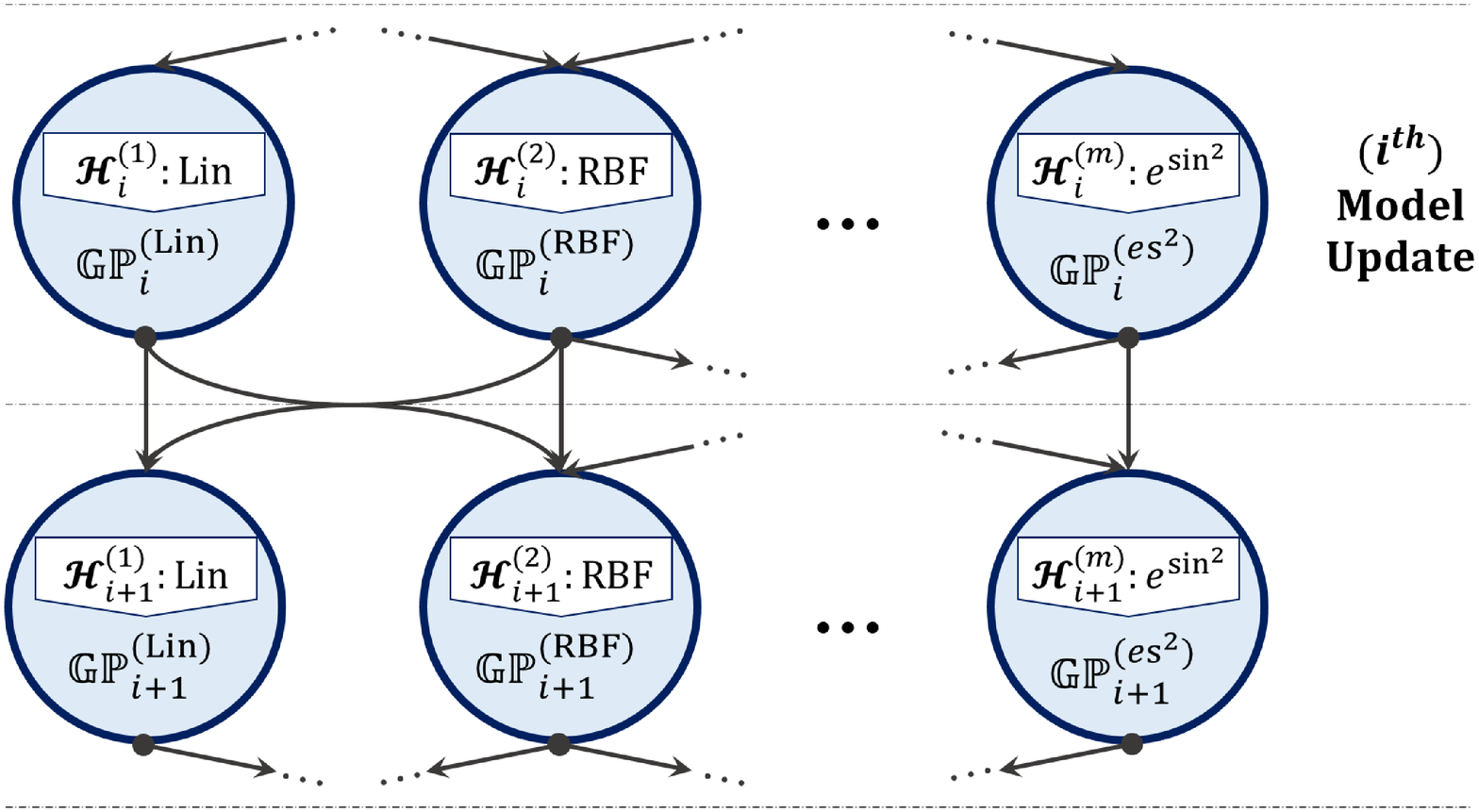}
\caption{Gaussian Process-based hybrid model update scheme}
\label{gp_update}
\end{figure}

In our settings, GPS latitude, longitude and elevation have been converted into ENU co-ordinations, then X-ENU and Y-ENU are treated as two separate time-series which should be learned from their own histories. Training window size has been set to 10 latest equally spaced received GPS samples in time (last 1 second) and a compound kernel type, composed of a linear and an RBF kernel, is selected due to our observations. Four different position tracking error thresholds,  i.e. 20 cm, 30 cm, 40 cm, and 50 cm, are investigated in this work. These values cover the range between minimum and maximum thresholds specified by SAE J2945\textbackslash 1, i.e 20 and 50 cm, respectively. The schematic representation of the proposed hybrid model communication policy is presented in Figure ~\ref{gp_update} and the pseudo-code of our algorithm is illustrated in Algorithm ~\ref{mbc:algo}. The evaluation results for the proposed framework are presented in the next section.

\begin{algorithm}
    \begin{algorithmic}
        \caption{Model-based Communication Algorithm} \label{mbc:algo}
        \REQUIRE Read CAN-bus at time $i$: $\mathcal{S}_i=\{x_{1,i}, ...,x_{n,i}\}$\\
        \STATE $T_0\gets T_{start}$ ; $i = 0$\\
        \WHILE{$T_0<T_{end}$}
        {
            \WHILE{$(PTE_{min}<th)$ or $(i=0)$ }
            {
                \begin{varwidth}[t]{\linewidth}
                    \STATE $i\gets i+1$;\par
                    \STATE $T_{next}\gets T_0+i$ \par
                    \FOR{$kernels \in \mathbfcal{H}_i^{T{next}}$}
                        \STATE $PTE_i$ = getPTE(\textit{kernel}, $\mathcal{S}_i$);
                    \ENDFOR \par
                    \STATE $PTE_{min} =\textbf{min}(PTE_i)$
                \end{varwidth}
            }
            \ENDWHILE 
            \par
            
            \begin{varwidth}[t]{\linewidth}
                \STATE \textbf{update} $T_0;$ \par
                \STATE \textbf{update} $\mathcal{S}_i;$ \par
                \STATE $PTE_{min} \gets \infty, i \gets 0;$ 
            \end{varwidth}
        }
        \ENDWHILE
    \end{algorithmic}
\end{algorithm}

\section{Evaluation}
System level performance gain of the MBC architecture stems from its two core components; model-based information exchange scheme and error-driven message transmit policy. In this section we first evaluate the performance gain originating from the communication policy in terms of offered channel load. Tracking accuracy of MBC is then compared against an error-driven raw information (conventional BSM) exchange policy (baseline) to further demonstrate its efficacy.

\begin{figure}[t]
\centering
\includegraphics[width=.48\textwidth]{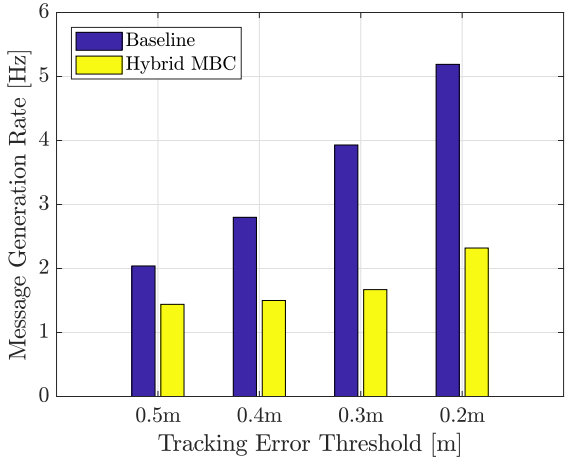}
\caption{Message scheduling rate for different choices of tracking error thresholds. For MBC, the total effective rate is shown here by summing the model update rate and the sub-model identification rate.}
\label{fig:rate-comparison}
\end{figure}

Since the error-driven model update is an integral part of the MBC, we evaluated the same message scheduling policy for our baseline sensor-generated data exchange scheme which uses constant-speed for position estimation. As mentioned earlier, the message scheduling rate in error-driven exchange depends on the selected tracking error threshold. This threshold may vary depending on the application requirements. We experimented with four different thresholds to determine the resulting message generation rate for MBC and baseline, as illustrated in Figure~\ref{fig:rate-comparison}. As the tracking error threshold gets stricter the message generation rate increases for both baseline and MBC. However, as the rate is significantly lower in MBC, it can accommodate higher number of transmitting entities compared to baseline, assuming over-the-air packet lengths of MBC and baseline are similar. Moreover, MBC experiences lower rate of packet collision comparing to its baseline counterpart in different traffic densities.

Now that the efficacy of MBC is established in terms of offered channel load, we seek to determine its tracking performance gain. Tracking performance of the MBC architecture is evaluated for the above mentioned vehicle trip which is carefully selected from the SPMD data-set \cite{SPMD:Data} on the merit of maneuver counts over the entire trip duration. To ensure fair comparison, message generation rate in baseline is chosen to be equal to the average model update rate computed in MBC. Tracking accuracy is determined in terms of position tracking error (PTE) which is defined as the 2D Euclidean distance between the actual and estimated vehicle position. Actual position at a given time instant is obtained from GPS logs and position estimation is calculated from received message information. Since PTE sampling is dependent on the availability of actual position updates, it can be done at most at the sampling rate of GPS updates, which is 10 Hz for SPMD dataset. For position estimation, MBC uses the most recent model parameters and associated model relevancy updates. In contrast, baseline vehicle position estimates are computed by coasting a vehicle's last received position update from the BSM to the error sampling instant (i.e., the GPS update instant), using a constant velocity mobility model.

We measured tracking errors for different packet error ratio (PER) levels. PER is indicative of the communication channel quality and is defined as the ratio of the missed packets to the transmitted packets. The PER metric can be interpreted from different perspectives. For a given traffic density, PER is typically an increasing function of sender-receiver separation distance. Conversely, for a given sender-receiver range, PER is an increasing function of traffic density. The \textit{``range"} interpretation is useful for assessing tracking performance at different ranges, while the \textit{``density"} interpretation is useful to evaluate range-specific system performance in different driving scenarios such as freeway with peak and off-peak hour traffic. Another way to interpret PER is based on line-of-sight conditions of sender-receiver pairs where links with dominant LOS results in low PER. This interpretation is applicable for performance evaluation in different driving environments such freeway and urban intersections.

\begin{figure}[t]
\centering
\includegraphics[width=.48\textwidth]{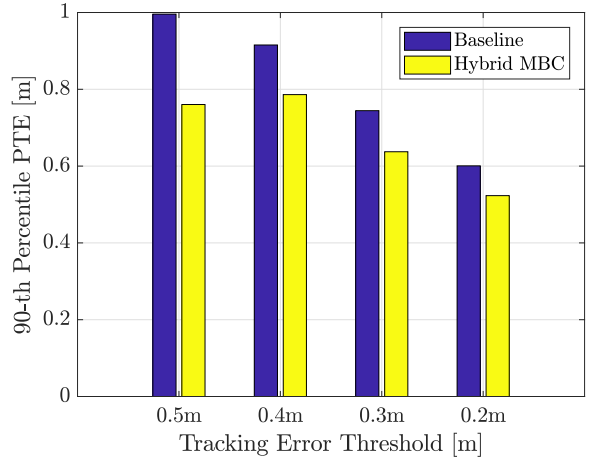}
\caption{Tracking error comparison for packet loss ratio of 40\%. Stricter (smaller) tracking error threshold translates to higher message transmit rate which eventually helps lowering the tracking error.}
\label{fig:te-ecdf-per}
\end{figure}


90-th percentile position tracking error (PTE), shown in Fig~\ref{fig:te-ecdf-per}, evidently suggests that MBC predicts positions more accurately than baseline. The higher tracking accuracy of MBC can be attributed to its capability of capturing higher order vehicle dynamics resulting from hard brakes and lane change maneuvers.



\section{Concluding Remarks}

The notable differences in tracking accuracy of different driving maneuvers, resulted from different modeling schemes motivates us to incorporate more complex model structures in comparison with what is the current state-of-the-art in vehicular society. More specifically, non-parametric Bayesian methods with different kernels, which are capable of being adapted to different maneuvers are potentially promising candidates for this purpose. Therefore, in this work we have proposed a Hybrid GP-based modeling scheme in combination with an error-driven model communication policy and investigated its performance against the same error-driven method of raw-information dissemination. A notable improvement is observed using our scheme against the base-line method through reduction of the required communication load as well as better tracking precision.

\bibliography{0Syscon2018-main.bib}{}
\bibliographystyle{unsrt}
\end{document}